\documentclass[]{article}

\usepackage[utf8]{inputenc}
\usepackage{amsmath}
\usepackage{graphicx}
\usepackage{color,soul}
\usepackage[margin=20mm]{geometry}
\usepackage{authblk}

\usepackage{booktabs}

\usepackage{dblfloatfix}
\usepackage[superscript,nomove]{cite}
\makeatletter
\renewcommand\@biblabel[1]{#1.}
\makeatother

\usepackage[normalem]{ulem}

\newcommand{\Arf}{A_{\rm rf}}

\newcommand{\phirf}{\phi_{\rm rf}}

\newcommand{\Athz}{A_{\rm THz}}
\newcommand{\tthz}{t_{\rm THz}}

\begin{document}
\title{Terahertz control of relativistic electron beams for femtosecond bunching and laser-synchronized temporal locking}

\author[1,2]{Morgan T. Hibberd}
\author[1,2]{Christopher T. Shaw}
\author[1,2]{Joseph T. Bradbury}
\author[1,3]{Daniel S. Lake}
\author[1,3]{Connor D. W. Mosley}
\author[1,3]{Sergey S. Siaber}
\author[1,5]{Laurence J. R. Nix}
\author[1,2]{Beatriz Higuera-Gonz\'{a}lez}
\author[1,4]{Thomas H. Pacey}
\author[1,4]{James K. Jones}
\author[1,4]{David A. Walsh}
\author[1,2]{Robert B. Appleby}
\author[1,5]{Graeme Burt}
\author[1,2]{Darren M. Graham}
\author[1,3]{Steven P. Jamison\thanks{s.jamison@lancaster.ac.uk}}

\affil[1]{The Cockcroft Institute, Sci-Tech Daresbury, Keckwick Lane, Warrington WA4 4AD, UK.}
\affil[2]{Department of Physics and Astronomy \& Photon Science Institute, The University of Manchester, Oxford Road, Manchester M13 9PL, UK.}
\affil[3]{Department of Physics, Lancaster University, Bailrigg, Lancaster LA1 4YB, UK.}
\affil[4]{Accelerator Science and Technology Centre, Science and Technology Facilities Council, Sci-Tech Daresbury, Keckwick Lane, Warrington WA4 4AD, UK.}
\affil[5]{School of Engineering, Lancaster University, Bailrigg, Lancaster LA1 4YW, UK.}

%\date{\today, \xxivtime }
\twocolumn[
  \begin{@twocolumnfalse}
    \maketitle
    \thispagestyle{empty} 
{ %\bf

Femtosecond relativistic electron bunches and micro-bunch trains synchronised with femtosecond precision to external laser sources are widely sought for next-generation accelerator and photonic technologies, from extreme UV and X-ray light sources for materials science, to ultrafast electron diffraction and future high-energy physics colliders. While few-femtosecond bunches have been demonstrated, achieving the control, stability and femtosecond-level laser synchronisation remains critically out of reach. Here we demonstrate a concept for laser-driven compression of high-energy (35.5\,MeV) electron bunches with temporal synchronisation to a high-power (few-TW) laser system. Laser-generated multi-cycle terahertz (THz) pulses drive periodic electron energy modulation, enabling subsequent magnetic compression capable of generating tuneable picosecond-spaced bunch trains with 30\,pC total charge and 50\,A peak currents, or to compress a single bunch by a factor of 27 down to 15\,fs duration. The THz-driven compression simultaneously drives temporal-locking of the bunch to the THz drive laser, providing a route to femtosecond-level synchronisation, overcoming the timing jitter inherent to radio-frequency accelerators and high-power laser systems. This THz technique offers compact and flexible bunch control with unprecedented temporal synchronisation, opening a pathway to unlock new capabilities for free electron lasers, ultrafast electron diffraction and novel plasma accelerators.
} 
    \\
  \end{@twocolumnfalse}]

%\Footer{\currfilebase }{\xxivtime}
%\setvruler[][][][3][][8mm][8mm][-2.5mm][]

The ability to deliver relativistic electron bunches with ultrashort bunch lengths and precise arrival timing locked to an external laser is a critical requirement as advanced accelerators move towards the femtosecond regime and below, unlocking a potential array of novel laser-based applications. These include examples such as pump-probe experiments using free-electron laser radiation \cite{Guo2024}, single-shot ultrafast electron diffraction \cite{Salen2025}, high-energy electron-laser collisions as direct probes of strong-field quantum electrodynamics \cite{Mirzaie2024} and for enabling future compact laser-plasma wakefield accelerators by exploiting synchronous external electron injection schemes \cite{Wu2021}, all highlighting the ever-growing fusion of accelerator and photon science.

Radio-frequency (RF) guns producing few-MeV electron beams have demonstrated sub-10 fs duration electron bunches, with 
an arrival-time jitter typically 50-100\,fs arising from fluctuations in the RF phase and amplitude\cite{Maxson2017,Zhao2018_2,Kim2020,Qi2020}. Taking advantage of the ability to use a single RF and laser source for few-MeV beams, 
 recent work has achieved sub-5\,fs bunch lengths and arrival-time jitter \cite{Yang2025}. For higher energy beams, the requirement for multiple RF systems (e.g. electron gun, linear accelerating cavities) and the two-way interplay between energy errors and timing errors makes laser-electron synchronisation more challenging. 
 For the higher energy beams, achieving the electron beam chirp for required femtosecond compression also requires an increased diversion of on-crest RF acceleration to off-crest generation of chirp. 
 Such an interplay of jitter sources and compression requirements can be seen in, for example, Pompili et al. \cite{Pompili2016}, with a demonstration of electron-to-laser timing down to 20\,fs 
with compression to 90\,fs bunch duration. 
In an advanced scheme using superconducting RF accelerators running at MHz repetition rates, fast intra-bunch train feedback has demonstrated 5\,fs level electron-to-laser synchronisation for subsequent micro-bunches but requiring the less-well synchronised bunches earlier in the train to be discarded \cite{Lautenschlager2021}. 
Alongside the highly developed RF acceleration technologies, single-stage laser-plasma wakefield  accelerators (LPWFA) offer high-energy electrons synchronised, at source, to the drive-laser.
Of note, recently combined LPFWA and passive RF correction has demonstrated energy stabilisation of laser-plasma generated electron bunches \cite{Winkler2025}. While this is a significant advance towards the exploitation of laser-plasma technology by reducing the energy spread and energy jitter by more than an order of magnitude, the stabilisation process converts the energy jitter to an additional picosecond-level temporal jitter on the electron beam, highlighting the challenges that remain. 
Future laser-plasma wakefield accelerators will require femtosecond-level synchronisation between high-power terawatt and petawatt lasers and electron beams for multi-stage acceleration, necessary for achieving the highest energy beams of particle physics ambitions. External injection of high-quality pre-prepared electron bunches has the potential to address challenges in electron-beam quality  when initial electrons are `self injected' from the background plasma, yet achieving the laser-electron bunch synchronisation and bunch compression required has remained stubbornly intractable for well over two decades.  
In particle-driven plasma wakefield acceleration, precisely controlled bunch pairs, or ideally bunch trains, with picosecond separation are required for drive-witness pairs or to resonantly enhance the acceleration wakefield. However, the requirement of picosecond spacing, or THz repetition rates, is mismatched to the nanosecond periods of RF-driven accelerators.

Ultimately, in both RF, plasma, and hybrid RF-plasma accelerators unavoidable jitter sources and RF constraints fundamentally limits the delivery of femtosecond-scale electron beams for a wide-range of sought-after applications. 
Laser-generated terahertz (THz) pulses offer a route to overcome these limitations, underlined by the huge progress achieved in THz-driven acceleration and manipulation of electron beams in recent years. Proof-of-principle demonstrations at relativistic beam energies include the first scalable THz linac \cite{Hibberd2020}, inverse free electron laser coupled energy modulation \cite{Curry2018}, staged acceleration showcasing high gradients \cite{Xu2021} and energy spread preservation \cite{Tang2021}, and a 
self-injection scheme
achieving MeV-scale energy gain \cite{Yu2023}. There is ongoing effort to drive low-energy electrons towards the relativistic regime \cite{Nanni2015,Walsh2017,Zhang2018,Zhang2019,Zhang2020,Nix2024} for compact X-ray light sources, development of THz-driven photoguns \cite{Huang2016} for precision diffraction and microscopy \cite{Ying2024} and THz-driven streaking of low-energy \cite{Kealhofer2016,Zhang2018} and relativistic \cite{Zhao2018,Zhao2019,Li2019,Wang2022} electron beams, providing new longitudinal diagnostic tools to characterise ultrashort bunches with femtosecond-scale resolution. 
Combining THz pulses with these low-energy beams, bunch compression and arrival time jitter suppression has been achieved\cite{Snively2020,Zhao2020}.

Despite the remarkable advances described above, femtosecond 
control of electron beams at the high energy and charge demanded by the next generation of advanced accelerator and laser-based applications remains a major challenge.
Here we demonstrate a concept exploiting laser-generated THz pulses for femtosecond control of fully relativistic 35.5\,MeV electron beams with bunch charges up to 30\,pC. Through THz-driven interaction mediated by a dielectric-lined waveguide structure, we experimentally achieve ultrafast energy modulation of electron bunches capable of compression into THz-frequency bunch trains, or single isolated bunches down to 15\,fs duration. As a consequence of the THz carrier phase being locked to the drive laser pulse envelope arrival time and the bunch compression process, we show the compressed electron bunches can be temporally locked to the drive laser with 25 fs rms arrival-time jitter despite the presence of 200 fs laser timing jitter. 
This temporal-locking is passive, or self-correcting, in that the electron bunch arrival time will be synchronised to and track the laser arrival time at a level more precise than the stabilisation of the laser system to the injected electron beam or the accelerator systems, opening a route to extremely tight and otherwise out-of-reach femtosecond synchronisation of high-energy electron beams to high-power laser systems.

\subsection*{Concept and implementation}

\begin{figure*}[htb!]
\includegraphics[width=\textwidth,keepaspectratio]{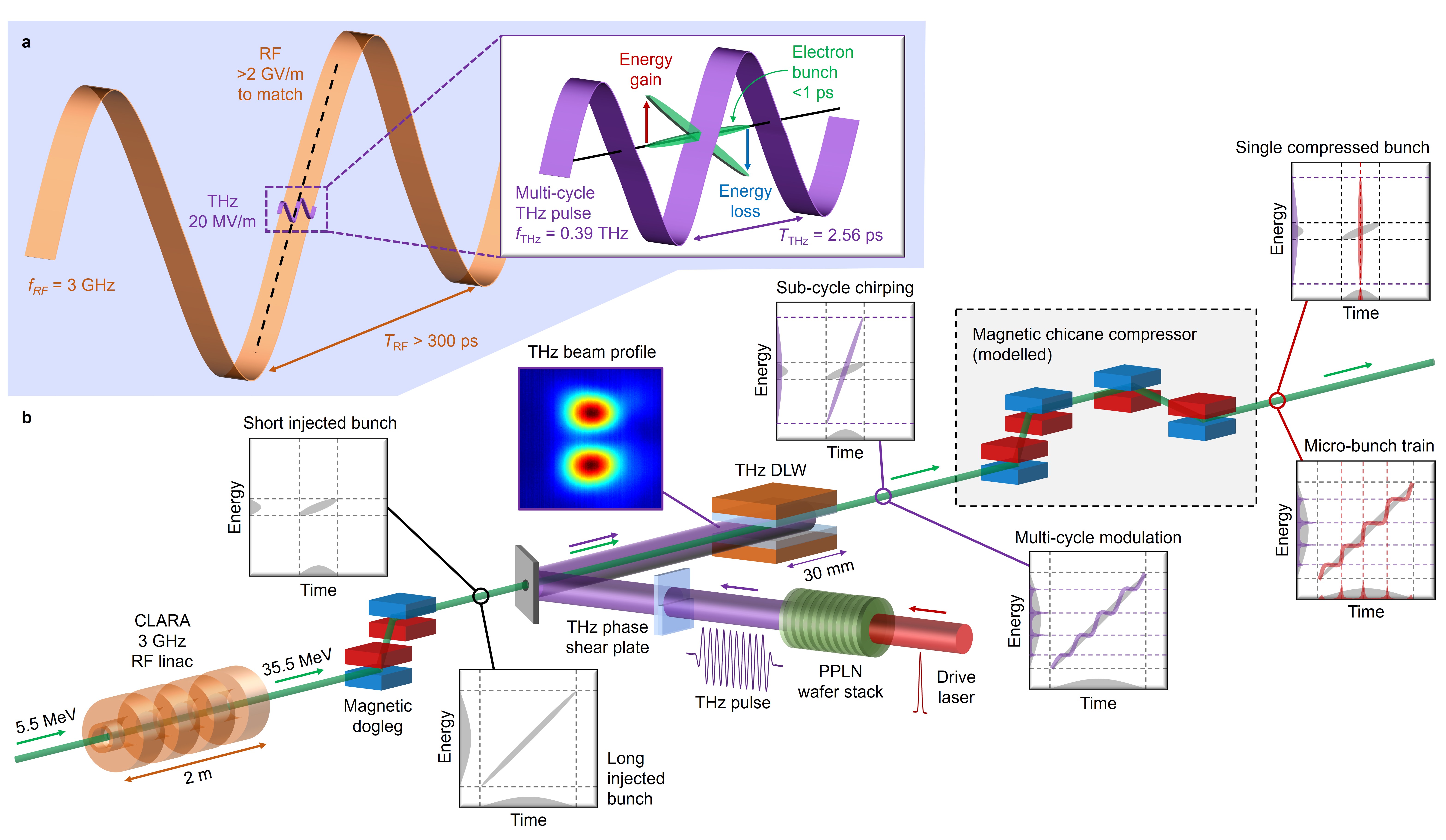}
\caption{\textbf{Concept for bunching and temporal locking.} \textbf{a} Schematic diagram demonstrating the extreme slope in accelerating gradient of high-frequency 0.39\,THz fields compared to few-GHz RF fields. Inset showing THz-driven chirping of an electron bunch injected at zero-crossing phase. \textbf{b} Overview of the concept showing the 35.5\,MeV electron bunches from the CLARA RF linac injected into the dielectric-lined waveguide (DLW) for THz-driven chirping/modulation and subsequent compression/bunching using a modelled magnetic chicane. The THz beam profile imaged with a THz camera at the DLW coupler entrance shows the quasi-TEM$_{01}$ mode used to excite the LSM$_{11}$ accelerating mode of the DLW. The representative time-energy density distributions of the electron bunch at each stage are given for both short (upper panels) and long (lower panels) injected electron bunches. }
\label{fig:concept}
\end{figure*}

THz pulses with energies exceeding $100\,\mu$J were generated from difference frequency mixing down-conversion of ultrafast near-infrared laser pulses from a Ti:sapphire terawatt (TW) laser system. The non-linear medium was large-area lithium niobate wafers with near-infrared anti-reflection coatings arranged in a periodically-poled configuration. Phase matching in the $\chi^{(2)}$-poled wafer structure leads to quasi-monochromatic THz pulses with a top-hat temporal envelope and a central frequency of around 0.39\,THz \cite{Mosley2023,Dalton2024}. 
The THz pulses are used to drive temporal compression of a relativistic electron bunch by driving acceleration and deceleration within a relativistic electron bunch. 
The THz-driven acceleration imprints an energy chirp with higher energy at the bunch tail than the head, which will result in compression after a subsequent drift (velocity-bunching) or dispersion (magnetic compression), as depicted in Fig.\,\ref{fig:concept}. To provide an extended interaction, propagation of the THz pulses is maintained in the confines of a dielectric-lined waveguide structure; the THz mode phase-velocity is matched to the relativistic electron beam velocity, $v_e = c\,(1 -  10^{-4})$. The THz pulse is converted to a quasi-TM$_{01}$ in free-space, generating the longitudinal electric field necessary for the particle acceleration interaction. A peak THz acceleration gradient of 
16\,MeV\,m$^{-1}$, comparable to that of conventional RF accelerator systems, is sustained over an interaction length of 8\,mm.
A resulting THz-driven chirp (time-energy correlation) is imposed on relativistic electron bunches (35.5\,MeV) delivered by the CLARA linear accelerator (linac) at Daresbury Laboratory \cite{Angal-Kalinin2020}. 
With a centre frequency of 0.39\,THz we obtain a chirp that would, if delivered over this compact interaction length, require an unfeasible peak acceleration gradient exceeding 2\,GeV\,m$^{-1}$ from the 3\,GHz RF system. Alternatively, regardless of interaction length, achieving this chirp from the RF linac would require the equivalent of at least 20\,MeV acceleration capability, greater than $50\%$ of the final beam energy, to be diverted to providing the chirp. 
In addition to single-bunch compression, the recurrence of the imposed chirp at each cycle of the narrowband THz pulse (2.56\,ps period), also provides the unique capability to generate multiple compressed micro-bunches in a train with ps-scale separation, as indicated schematically by the time-energy charge density distributions in Fig.\,\ref{fig:concept}b. Furthermore, with the imposed chirp directly synchronised to the laser generating the  THz pulses, an active temporal-locking between compressed electron bunches and the drive laser can be achieved. We show that the combination of high-gradient and high-frequency allows the temporal-locking to dominate and suppress the typical and significant bunch timing jitter arising from the RF accelerator and laser systems \cite{Pompili2016,Angal-Kalinin2020}. 

We have experimentally demonstrated these concepts using long ($\sigma_t$\,=\,2.5\,ps rms) and short, ($\sigma_t$\,=\,400\,fs rms) 35.5\,MeV electron bunches at the CLARA accelerator. The THz source was provided by laser-driven large-area periodically-poled lithium niobate (PPLN) wafer stacks \cite{Mosley2023}, delivering multi-cycle THz pulses with top-hat temporal profile up to 10 cycles long and pulse energies (at source) of up to 100\,$\mu$J. The THz pulses were transported in free-space via off-axis parabolic mirrors and coupled into a dielectric-lined rectangular waveguide (DLW) supporting a phase-velocity matched LSM$_{11}$ mode at 0.39\,THz\,~\cite{Hibberd2020}, providing on-axis longitudinal fields for acceleration. To achieve coupling to the LSM$_{11}$ waveguide mode, the $\text{TEM}_{00}$ mode generated at source was converted to a quasi-$\text{TEM}_{01}$ mode using a $\lambda/2$ phase shear plate in the free-space transport region, as shown in Fig.\,\ref{fig:concept}b. Following the interaction, measurements of the electron energy spectrum as a function of THz injection phase (controlled with femtosecond precision by an optical delay stage) and THz pulse energy were recorded for each bunch configuration. 
The subsequent bunch compression of a chicane was modelled through applying a first-order temporal dispersion to the as-measured electron-beam time-energy distribution. 

\subsection*{THz-frequency bunch trains}

\begin{figure*}[hb!]
\includegraphics[width=\textwidth,keepaspectratio]{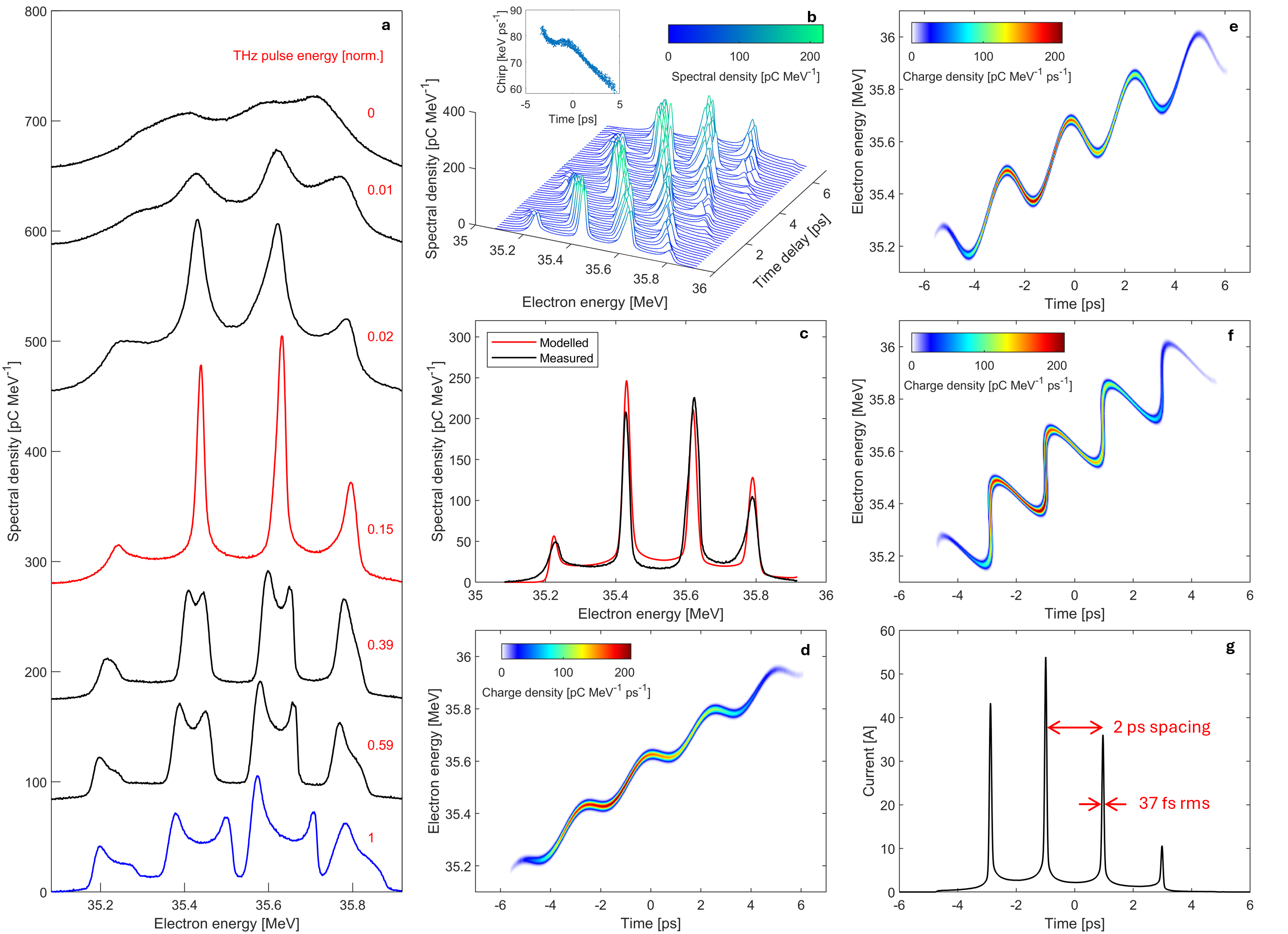}
\caption{\textbf{Multi-cycle energy modulation for micro-bunch trains.} \textbf{a} The measured THz-modulated electron energy spectra for varying THz pulse energy using approximately linearly-chirped 2.5\,ps\,rms injected bunches. The narrow-peak splitting threshold at 15\% of the maximum THz energy is highlighted in red, with the corresponding \textbf{b} measured time-delay scan (inset showing the extracted bunch chirp), \textbf{c} modelled fitting to the modulated energy spectrum and \textbf{d} full time-energy density distribution. For the highest THz pulse energy (maximum peak-splitting modulation highlighted in blue), the corresponding time-energy density distributions \textbf{e} before and \textbf{f} after compression, with the \textbf{g} predicted time projection of the micro-bunch train using an applied temporal dispersion of $D=$\,3.1\,ps\,MeV$^{-1}$.}  
\label{fig:bunchtrains}
\end{figure*}

The multi-cycle nature of our THz drive pulses provide the ability to directly manipulate electron bunches periodically at THz frequencies, which is not possible using conventional RF fields, and we explored this potential for the generation of trains of ultrashort micro-bunches with picosecond-scale spacing. For this experiment, the CLARA accelerator was configured to provide approximately linearly-chirped 2.5\,ps rms duration electron bunches with 30\,pC charge, and energy spread of 450\,keV FWHM. THz-driven multi-cycle energy modulation was observed on a single-shot basis, as a function of THz pulse energy (Fig.\,\ref{fig:bunchtrains}a) and timing (Fig.\,\ref{fig:bunchtrains}b). Typically 50 spectra were recorded for each configuration and time delay, revealing the shot-to-shot variation in the injected and THz modulated electron bunches. In Fig.\,\ref{fig:bunchtrains}a, as the THz pulse energy was increased, an initially sinusoidal energy modulation evolved into narrow spectral features, with the narrowest energy peaks occurring when the localised THz-induced negative chirp cancelled out the positive linear chirp of the injected bunch. Further increase in the THz energy led to a periodic reversal of chirp and splitting in the spectral peaks. Observation of the splitting threshold, achieved here due to the 10-fold increase in THz-induced energy gain over previous experiments \cite{Hibberd2020}, allowed for a model-free calibration of the peak THz-induced energy gain. Firstly, the bunch length and chirp were determined by correlating the periodic energy peaks with the known THz period as a function of relative THz-electron injection timing \cite{Hibberd2020}. The experimental timing scan is shown in Fig.\,\ref{fig:bunchtrains}b, with the measured chirp of the injected bunch in the inset. A dominant linear contribution to the chirp of approximately 70\,keV\,ps$^{-1}$ was determined, with the observed temporal variations revealing the presence of higher-order chirp components. With knowledge of the underlying chirp, the splitting threshold can then be identified with the formation of a "staircase" charge density distribution as shown in Fig.~\ref{fig:bunchtrains}d. From this analysis, a peak THz-driven energy gain of 104\,keV was determined for these specific experiments (a higher gain was obtained in the single-bunch compression experiments discussed later). With the independent calibration of the THz-driven energy gain, the width of the narrow spectral peaks (see Fig.~\ref{fig:bunchtrains}c) provides a direct measurement of the energy spread at a given temporal position within the bunch known as the time-slice energy spread, a critical beam parameter that sets the lower bound on the achievable compressed bunch duration. Here, a rms slice energy spread of $\sigma_U=8\,$keV was determined for this long-bunch configuration. Through these measurements the complete time-energy density distribution of the THz-modulated electron bunch is determined, as shown in Fig.\,\ref{fig:bunchtrains}d. 

From our full characterisation of the injected bunch using lower THz pulse energies (approximately 15\% of the maximum available), we explored the bunch compression capability for high-energy THz-driven interactions. At the maximum THz energy used for this bunch configuration, we obtain a highly over-split electron spectrum, as shown in Fig.\ref{fig:bunchtrains}a. The corresponding time-energy density distribution given in Fig.\,\ref{fig:bunchtrains}e demonstrated periodic localised THz-induced chirping up to a total of 0.325\,MeV\,ps$^{-1}$ repeating every 2.56\,ps along the bunch. This known energy modulated bunch distribution was projected through a matched magnetic chicane with an absolute temporal dispersion of \mbox{$D$ = -3.1\,ps\,MeV$^{-1}$} to obtain the charge density distribution in Fig.\ref{fig:bunchtrains}f and corresponding temporal projection of the micro-bunch train with period of approximately 2\,ps, as shown in Fig.\ref{fig:bunchtrains}g. The micro-bunch period differs from the THz frequency through the temporal compression associated with the underlying linear chirp of the injected electron bunch, which offers tuneability to the micro-bunch spacing. The observed THz-induced periodic chirp would enable the compression of each micro-bunch to a duration of 37\,fs rms, containing a peak charge of 5\,pC with peak currents of up to 50\,A in each micro-bunch. Residual charge from the 30\,pC injected bunch outside the current peaks, formed an underlying low-current pedestal. 
Extrapolation from the experimental results indicates that with realistic and achievable reduction of injected time-slice energy spread and an increase in the THz accelerating gradient, micro-bunches with peak currents approaching the kA regime can be achieved, taking the bunches into the regime typical of free-electron lasers.
%\hl{[JB: changed 50 MeV/m to 100 MeV/m. The calculation assumed a fivefold increase in accelerating field strength. Other details: micro-bunch duration 8.9 fs FWHM (3.8 fs rms), complete charge transmission through DLW assumed, no space-charge.]}

\subsection*{Femtosecond bunch compression}

\begin{figure*}[tb!]
\includegraphics[width=\textwidth,keepaspectratio]{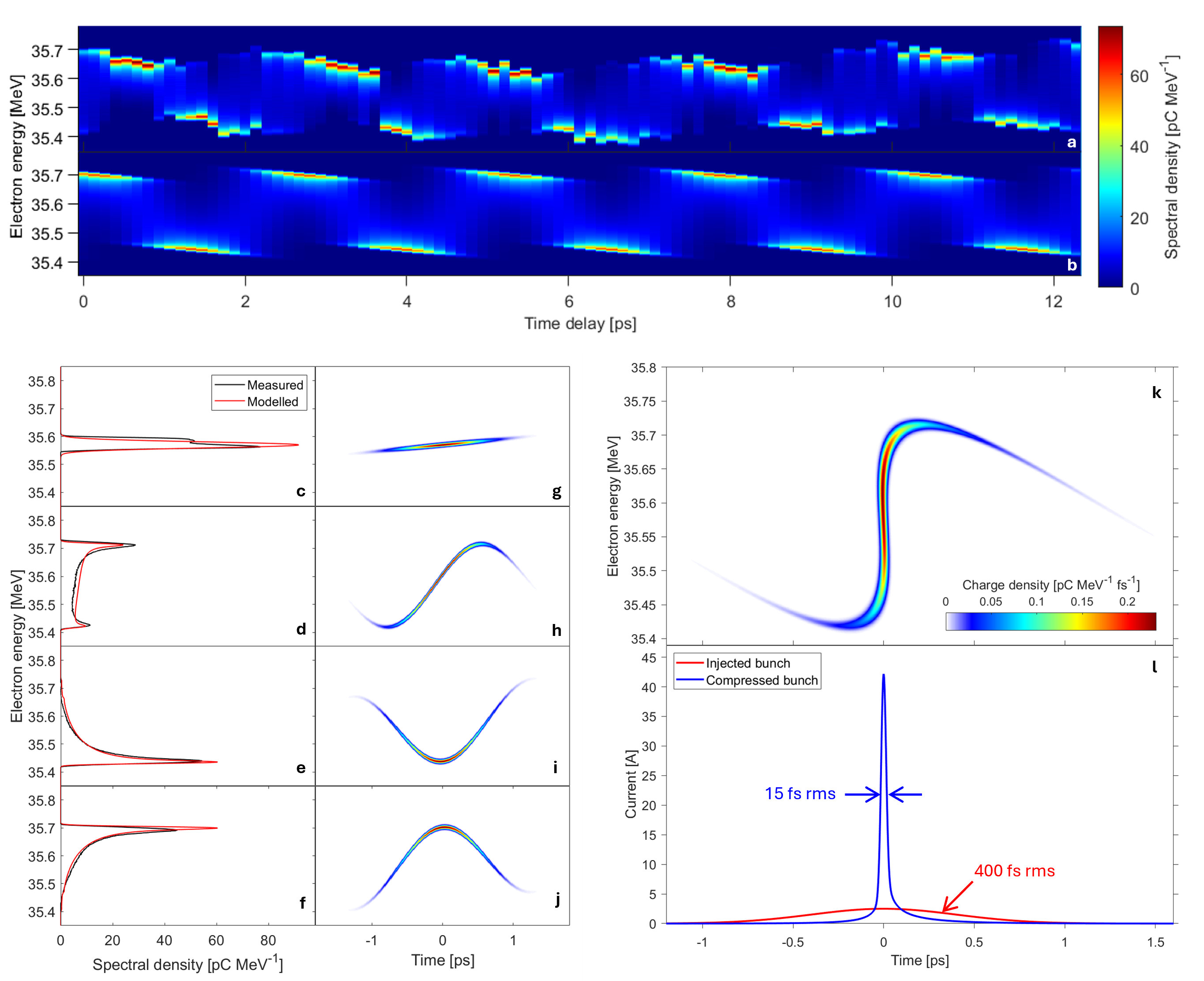}
\caption{\textbf{Sub-cycle THz interactions for single-bunch compression.} For 400\,fs rms injected bunches, the \textbf{a} measured and \textbf{b} modelled energy spectra as a function of THz-electron injection time. Selected spectral projections \textbf{c} without THz interaction and with THz \textbf{d} zero-crossing, \textbf{e} decelerating and \textbf{f} accelerating phase, with \textbf{g-j} the corresponding modelled time-energy density distributions. From the zero-crossing phase with maximum THz-induced chirp, the predicted \textbf{k} time-energy density distribution and \textbf{l} temporal projection of the electron bunch following optimised magnetic compression using a matched temporal dispersion of $D=$\,-3.01\,ps\,MeV$^{-1}$.} 
\label{fig:singlebunch}
\end{figure*}

To demonstrate strong compression of isolated electron bunches, the CLARA accelerator was configured to provide sub-ps low energy spread bunches (duration \,400\,fs,  total energy spread of 50\,keV FWHM, and 2\,pC charge) for injection within a half-cycle of the THz field. To maintain low projected energy spread, the RF linac was operated near the peak of the RF accelerating field, providing low-chirp electron bunches to the THz interaction point. The electron energy spectrum was observed as a function of THz-electron timing over many THz cycles, revealing the periodic interaction shown in Fig.\ref{fig:singlebunch}a. For comparison with the experimental data, the electron spectra were modelled from an injected electron bunch with a simple Gaussian temporal profile with Gaussian time-slice energy spread and linear-only chirp, with the optimised model timing scan shown in Fig.\ref{fig:singlebunch}b. Selected individual spectra and the corresponding time-energy density distributions are shown in Fig.\,\ref{fig:singlebunch}c-f and Fig.\,\ref{fig:singlebunch}g-j, respectively, highlighting the key THz-electron interaction phases for maximum THz-driven chirping at the zero-crossing phase, and the maximum deceleration or acceleration at each peak of the THz field cycle. From the analysis, the injected bunch duration was determined to be 400\,fs rms with a residual linear chirp of 20\,keV\,ps$^{-1}$ and time-slice energy spread of 10\,keV\,FWHM. A THz pulse energy of up to $120\,\mu$J at source resulted in a peak energy gain of 133\,keV, measured directly from the spectral peak positions in \mbox{Fig.\,\ref{fig:singlebunch}c-f}. 
At the injection zero-crossing phase in Fig.\,\ref{fig:singlebunch}d and h, a maximum linear chirp of 0.345\,MeV\,ps$^{-1}$ was imposed on the bunch. Through a modelled chicane, this would require a dispersion of $D = -3.01$\,ps\,MeV$^{-1}$ for optimal compression, achieving a bunch length of 15\,fs rms, as shown by the predicted time-energy density distribution in Fig.\ref{fig:singlebunch}k and corresponding temporal projection in Fig.\ref{fig:singlebunch}l. This represents a compression factor of up to 27, over an order-of-magnitude higher than previous THz-driven compression schemes \cite{Snively2020,Zhao2020}.

\subsection*{Electron-to-laser temporal locking}

\begin{figure*}[tb!]
\centerline{
\includegraphics[width=17cm]{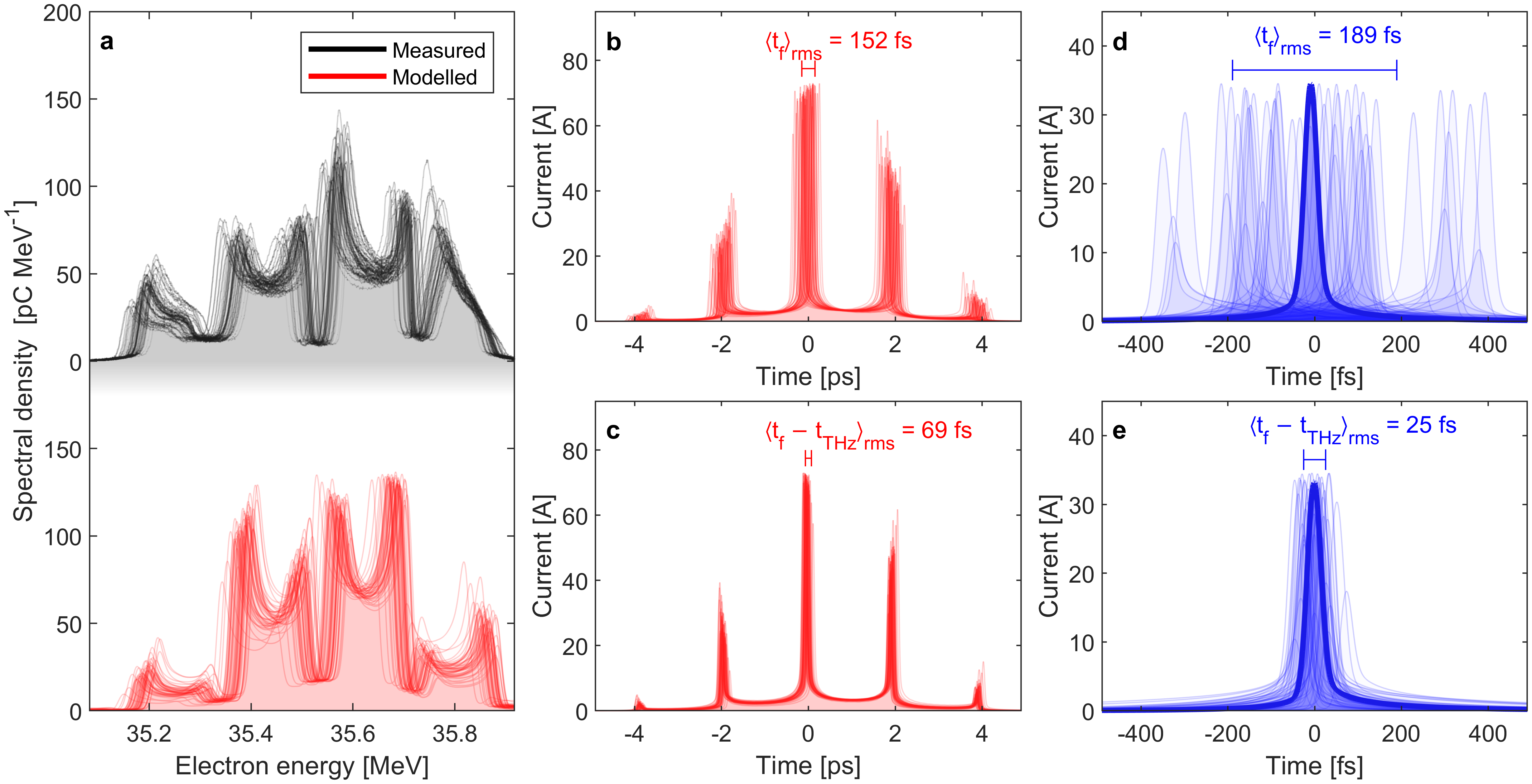}
}
\caption{\textbf{Laser-driven temporal locking of femtosecond bunch trains and single bunches.} \textbf{a,} Observed electron energy spectra (50 shots) using the CLARA long-bunch configuration with the highest energy THz modulation, and the corresponding modelled spectra including jitter in the accelerator and laser systems. \textbf{b,} Projected temporal charge distribution of the corresponding compressed bunch train with respect to an independent reference clock, and \textbf{c} with respect to the drive laser. \textbf{d,} Projected temporal charge distribution of the compressed sub-cycle single bunches with respect to an independent reference clock, and \textbf{e} with respect to the drive laser.}
\label{fig:jitter}
\end{figure*}

To provide a quantitative analysis of the temporal locking capability, we consider the jitter contributions from the accelerator configuration shown schematically in Fig.\,\ref{fig:concept}. We consider the evolution of the time-energy density distribution through the RF linac, the THz modulator and compression stages in the presence of jitter around set-points in the linac injection timing and energy; linac RF amplitude and phase, and THz amplitude and phase (timing) (see methods and supplementary information for further details). Defining electron bunch compression as occurring when the arrival time is independent of the initial injection time, we can determine the chicane temporal dispersion that is required for bunch compression. Of note, the required dispersion is inversely proportional to the chirping drive frequency, with the two-orders of magnitude increase in THz frequency over typical RF compression enabling a decrease in the temporal dispersion required for compression and reduced sensitivity of the arrival time on the mean electron energy.

To evaluate the compression and arrival-time locking, we consider a fixed dispersion matched to the nominal accelerator configuration, and the evolution of the time-energy density distribution in the presence of jitter about this nominal setting. We have carried out this evaluation for the THz-frequency bunch trains and the isolated ultrashort bunch configurations described above. The independently determined or estimated jitter sources are summarized in Table~\ref{jittersetpoints:tab} in the Methods. For the THz-frequency bunch trains, the measured and predicted electron energy spectra over 50 shots are given in Fig.\ref{fig:jitter}a, with the observed energy fluctuations in the measured spectra closely matching that obtained from particle tracking in the presence of jitter in the accelerator and laser systems. The corresponding bunch trains predicted after compression in the dispersive chicane are shown with respect to an independent stable reference clock (Fig.\ref{fig:jitter}b) and with respect to the THz pulse (and by definition the drive laser) arrival time (Fig.\ref{fig:jitter}c). A significant reduction in the arrival-time jitter in the bunch train is observed when referenced to the arrival timing of the THz drive laser, demonstrating temporal locking of the micro-bunches to the laser pulse.

For the single compressed bunch configuration, the equivalent 50 shots are again shown with respect to the timing of the independent reference clock (Fig.\ref{fig:jitter}d) and the THz drive laser (Fig.\ref{fig:jitter}e). In the latter case, we find an arrival time jitter of 25\,fs with respect to the THz source, almost an order of magnitude below the THz source jitter of 200\,fs (estimated from previous electron-to-laser timing measurements performed on the CLARA accelerator \cite{Walsh2023}). This level of temporal locking to the THz pulse is largely independent of the timing stability of the THz source, which is critically important for laser-electron beam experiments given the often high temporal jitter and drift present in high-energy terawatt femtosecond laser systems.

\subsection*{Conclusions and future potential}

The use of laser-generated THz pulses together with conventional RF accelerator technology offers a capability to control the temporal properties of relativistic electron beams on a femtosecond timescale. While modern RF accelerators provide unmatched spatial (emittance) and spectral (energy spread) beam quality, the low frequency and inherent RF-based jitter limit their ability to manipulate ultrashort high-charge bunches with femtosecond-scale temporal precision and laser synchronisation. We have demonstrated the capability to compress a fully relativistic beam with significant charge (30\,pC) into a THz frequency bunch-train that is readily tuneable in micro-bunch spacing, even with a fixed-frequency THz source. From our experimentally achieved THz-imposed bunch chirp, we have shown a route to compression of single electron bunches down to 15\,fs and below, and the potential to reach the kA peak currents typically required for free-electron laser applications. We have also demonstrated that the THz-driven bunching can provide an inherent and tight temporal locking between an external laser and the compressed electron beams, providing an opportunity to address the extreme synchronisation requirements for controlled external injection into advanced laser-plasma accelerators for future high-energy applications.

\clearpage

\pagebreak

\section*{Methods}

\textbf{Electron beam.} 
The Compact Linear Accelerator for Research and Applications (CLARA) test facility at Daresbury Laboratory provided relativistic electron bunches by photo-exciting a copper cathode with 266\,nm, 2\,ps FWHM laser pulses at a repetition rate of 10\,Hz, accelerating them to 5.5\,MeV in a 3\,GHz RF gun, and subsequently up to 35.5\,MeV in the RF linac stage. The electron bunches were then transported to an experimental area using a magnetic dogleg, which together with the RF linac, enabled the longitudinal phase space of the beam to be manipulated. The accelerator was configured to produce either short bunches with low energy spread (400\,fs rms, 2\,pC, 50\,keV FWHM) or long bunches with high energy spread (2.5\,ps rms, 30\,pC, 450\,keV FWHM). In the experimental area, quadrupole triplets were used to focus the electron beam into the THz-driven DLW and subsequently image it into the energy spectrometer.
\\
\\
\textbf{Laser system} The experiment was performed using a terawatt (TW) laser system that delivered up to 600\,mJ laser pulses with a centre wavelength of 800\,nm, transform-limited pulse duration of 50\,fs and repetition rate of 10\,Hz. The TW laser was synchronised to the CLARA photo-injector laser through a commercial phase-lock-loop electronics module (Synchrolock, Coherent). Following grating compression and beam transport to the CLARA experimental area, a maximum laser pulse energy of 350\,mJ was available for THz generation. 
\\
\\
\textbf{Terahertz generation and transport} Narrowband THz pulses centred at 0.39\,THz were generated by photo-exciting a periodically poled lithium niobate (PPLN) wafer-stack source \cite{Mosley2023} constructed from \textit{x}-cut single-crystal wafers of congruent lithium niobate (diameter of 50.8\,mm, thickness of 135\,$\mu$m, and 760-840\,nm anti-reflection coating on both sides). The periodic poling was achieved by inverting the direction of the \textit{z}-axis for adjacent wafers in the stack, with up to 20 wafers (10 cycles) used for experiments. For optimal THz generation, the TW laser pulse duration was slightly under-compressed to 750\,fs and variable focusing was used to achieve a pump fluence of approximately 500\,mJ\,cm$^{-2}$ at the PPLN source for a range of laser energies. The THz pulse energy was measured at the exit of the PPLN source using a pyroelectric detector (THZ-I-BNC, Gentec) with values up to 120\,$\mu$J and fluence-dependent conversion efficiencies ranging from 0.03-0.1\%. The THz radiation was routed by metallic mirrors into a vacuum chamber through a quartz window and focused with a 90 degree off-axis gold-coated parabolic mirror of focal length of 152.4\,mm into the coupler of the DLW. The free-space quasi-TEM$_{01}$ THz mode, required to drive the LSM$_{11}$ accelerating mode in the DLW, was generated using a 40\,mm-diameter PTFE phase-shear plate in the THz beam path, with a thickness difference between the top and bottom halves of approximately 800\,$\mu$m designed to produce a half-cycle phase-shift. Losses due to THz transport and mode-generation resulted in only approximately 20\% of the source energy arriving at the DLW coupler position. 
\\
\\
\textbf{Waveguide.} The dielectric-lined waveguide (DLW) consisted of a hollow, rectangular copper structure lined at the top and bottom with 70\,$\mu$m-thick fused quartz, leaving a 460\,$\mu$m-thick vacuum aperture of 30\,mm length and 1.2\,mm width for electron-beam propagation. A tapered horn structure at the entrance to the DLW aperture (dimensions of 3\,mm by 3\,mm and 23\,mm in length) aided coupling of the THz radiation into the accelerating mode of the DLW. A 45 degree aluminium mirror for THz reflection, with a 400\,$\mu$m aperture aligned to the DLW for electron beam transmission, permitted precise spatial alignment of the THz and electron beams through the DLW. 
\\
\\
\textbf{Compression and bunch train modelling.} The electron bunch was primarily modelled by simulating the time-energy density distributions at both DLW injection and after THz interaction. The bunch chirp and slice energy spread used in the modelling were obtained directly from THz phase scans using the narrow-peak splitting threshold in the multi-cycle modulation regime. Due to the use of multi-cycle THz pulses, superposition of a sinusoidal energy change to the time-energy density distributions was used to model the THz interaction. To model the bunch compression through a dispersive magnetic chicane, delay in electron arrival was accumulated by a linear amount proportional to dispersion $D$ and the energy difference from the reference energy of 35.5\,MeV. Linear particle tracking simulations through the CLARA dipole spectrometer beamline were employed for understanding of the spectrometer performance. 
\\
\\
\textbf{Charge density distribution evolution}
The propagation of an electron (or density distribution) with initial energy and time $U_i, t_i$, respectively, can be tracked through the accelerator, to a final electron energy and arrival time $U_f, t_f$,
\begin{equation*}
U_f = f(U_i,t_i,\{\alpha_k\}), \quad\quad 
 t_f = g(U_i,t_i,\{\alpha_k\})
     \label{Generaltracking:eqn}
\end{equation*}
where the functions $f, g$ can be found from cascading of the effects from individual accelerator elements, and $\alpha_{\rm k} \in \{\Arf,\phirf,\Athz,\tthz\}$ is a set of parameters describing the RF amplitude and phase, and THz amplitude and timing. From the symplectic nature of the electron trajectory, the charge density distribution $\rho(U,t)$ evolves such that the localised charge density distribution tracks through the accelerator system in the same manner as an individual particle,
\begin{eqnarray*}
&& \rho_f(U_f,t_f) = \rho_{\rm inj}(U_i(U_f,t_f,\{\alpha_k\}),t_i(U_f,t_f,\{\alpha_k\})) \nonumber \\
&& \quad  =  \rho_{\rm inj}(f^{-1}(U_f,t_f,\{\alpha_k\}), g^{-1}(U_f,t_f,\{\alpha_k\}))
\end{eqnarray*}
\\
\textbf{Temporal-locking and time-jitter modelling.}
To analyse the arrival time jitter requires quantitative understanding of the effects of jitter in the upstream accelerator systems on the electron bunch injected into the THz modulator. 
For the lattice of Fig.\ref{fig:concept} the transport functions were evaluated for the fixed setpoints and estimated rms jitter offsets \cite{Angal-Kalinin2020,Walsh2023} as given in Table \ref{jittersetpoints:tab}, and with a configuration-specific fixed temporal dispersion $D$ in the chicane (see main text).

\begin{table}[hbt]
\centering
  {\small 
\begin{tabular}{lllc}
 \toprule
  System \quad   & set-point  & rms jitter & comment     \\[1.2ex]
            & $\alpha_{k,\,{\rm ref}} \quad $      & 
               $\Delta\alpha_k$  &   \\ 
                \midrule
$U_{\rm rf}$    &    30\,MeV &    8\,keV &                         \\ 
$\phi_{\rm rf}$  &    0$^{o} $ & $10^{-3}\,{\rm rad.}$   &    single bunch              \\ 
                 &    -15$^{o} $ & $10^{-3}\,{\rm rad.}$   &   bunch trains              \\ 
                 &             &  $\,\,\simeq 53$\,fs  &                               \\ 
$U_{\rm THz}$    &      133\,keV  &    3\,keV&     single bunch                         \\ 
                   &      104\,keV  &    2\,keV&     bunch trains                         \\ 
$t_{\rm THz}$    &    0\,fs &   200\,fs&                         \\ 
$t_{\rm inj}$    &    0\,fs     &   200\,fs &                            \\ 
$U_{\rm inj}$    &      5.5\,MeV &   2\,keV  &                 \\            
 \bottomrule
\end{tabular}
\caption{Accelerator and laser set-point values for the single-bunch and bunch train configurations, and the corresponding rms jitters used for the demonstration of temporal-locking to the THz source.}
\label{jittersetpoints:tab}
}
\end{table}

\section*{Acknowledgements}
We wish to acknowledge the technical and scientific teams at the Compact Linear Accelerator for Research and Applications (CLARA) facility for their support and help with the operation of the accelerator. We also wish to acknowledge Peter G. Huggard and Mat Beardsley from Rutherford Appleton Laboratory (RAL)-Space for the manufacture of the dielectric-lined waveguide structure. \\This work was supported by the United Kingdom Science and Technology Facilities Council (STFC) [Grant Nos. ST/P002056/1, ST/V001612/1, ST/X004090/1], and STFC studentship awards to C. T. Shaw and J. T. Bradbury [Project Ref. 2488454, 2649517].

\section*{Author contributions}
All authors participated in the experiment and contributed to data analysis. C.T.S., J.T.B., R.B.A. and S.P.J. modelled the energy spectra and time-energy distributions for single-bunch compression. J.T.B., R.B.A. and S.P.J. modelled the energy modulations for micro-bunch trains and the temporal locking. C.D.W.M., M.T.H., D.S.L., D.A.W. and D.M.G. developed the laser setup and THz source. L.J.R.N and G.B. undertook analyses of the THz waveguide coupling. B.H.-G. and D.M.G characterised the DLW. T.H.P. and J.K.J. analysed the beam dynamics of the CLARA accelerator. M.T.H., D.M.G. and S.P.J. wrote the manuscript with contributions from all. R.B.A., G.B., D.M.G. and S.P.J. managed the project.

\section*{Additional information}
All correspondence should be addressed to S.P.J.

\end{document}